\documentclass[aps,prd,nofootinbib,superscriptaddress,twocolumn]{revtex4}

\usepackage{latexsym}
\usepackage{amssymb}
\usepackage{epsfig,amsmath,graphics}
\usepackage{multirow}

\newcommand{\gsim}{\lower.7ex\hbox{$\;\stackrel{\textstyle>}{\sim}\;$}}
\newcommand{\lsim}{\lower.7ex\hbox{$\;\stackrel{\textstyle<}{\sim}\;$}}

\newcommand{\OO}{\mathcal{O}}
\newcommand{\LL}{\mathcal{L}}

\newcommand{\fb}{\text{fb}}

\newcommand{\MeV}{\text{ MeV}}
\newcommand{\GeV}{\text{ GeV}}
\newcommand{\TeV}{\text{ TeV}}

\newcommand{\hc}{\text{ h.c.}}

\newcommand{\Br}{\text{ Br}}

\newcommand{\half}{\frac{1}{2}}

\newcommand{\DO}{{\mbox{DO\hspace{-0.105in}$\not$\hspace{0.11in}}}}

\newcommand{\MET}{\mbox{$E_T\hspace{-0.21in}\not\hspace{0.15in}$}}

\begin{document}

\pagestyle{plain}

\title{
\begin{flushright}
\mbox{\normalsize SLAC-PUB-13534}
\end{flushright}
\vskip 15 pt

Discovering the Higgs with Low Mass Muon Pairs}

\author{Mariangela Lisanti and Jay G. Wacker}
\affiliation{
SLAC, Stanford University, Menlo Park, CA 94025\\
Physics Department, Stanford University,
Stanford, CA 94305}

\date{\today}
\begin{abstract}
Many models of electroweak symmetry breaking have an additional light pseudoscalar.  If the Higgs boson can decay to a new pseudoscalar, LEP searches for the Higgs can be significantly altered and the Higgs can be as light as 86 GeV.  Discovering the Higgs boson in these models is challenging when the pseudoscalar is lighter than 10 GeV because it decays dominantly into tau leptons.  In this paper, we discuss discovering the Higgs in a subdominant decay mode where one of the pseudoscalars decays to a pair of muons.  This search allows for potential discovery of a cascade-decaying Higgs boson with the complete Tevatron data set or early data at the LHC.        
\end{abstract}

\maketitle

\section{Introduction}
The last unexplored frontier of the Standard Model is electroweak symmetry breaking, the process by which the Higgs field obtains a vacuum expectation value and gives mass to the $W^{\pm}$ and $Z^0$ gauge bosons.  One of the major goals of current colliders is to discover the Higgs boson and understand the dynamics that give rise to electroweak symmetry breaking.  There have been direct and indirect searches for the Standard Model (SM) Higgs at LEP and the Tevatron.  The current lower bound on the Higgs mass, 
\begin{equation}
m_{h^0} > 114.4 \text{ GeV}  \text{     (95\% confidence)}, \nonumber
\end{equation}
 comes from searches at LEP for $e^+ e^- \rightarrow Z^0 h^0$, with the SM Higgs decaying to a pair of taus or bottom quarks \cite{Barate:2003sz}.  Recently, combined Higgs searches from the CDF and $\DO$ experiments at the Tevatron excluded a SM-like Higgs of $169 \GeV\le m_{h^0}\le 171 \GeV$ \cite{Bernardi:2008ee}.  
 
While direct searches for the Higgs point towards a heavy mass, indirect bounds from electroweak constraints place a limit on how heavy the mass can be.  In particular, the best fit for a SM Higgs mass is 77 GeV with a 95\% upper bound of 167 GeV \cite{LEP:2008}.  This limit comes from measurements of electroweak parameters that depend logarithmically on the Higgs mass through radiative corrections.  There is tension between the direct and indirect measurements; only a narrow window of masses for the SM Higgs satisfies both results.      

On the theoretical side, a light Higgs is preferred within the Minimal Supersymmetric Standard Model (MSSM).  Requiring a natural theory and minimizing fine tuning drives the Higgs mass below the LEP direct bound.  In the MSSM, there are two new Higgs chiral superfields, $H_u$ and $H_d$, that result in two CP-even scalars $H^0$ and $h^0$, the CP-odd scalar $A^0$, and the charged Higgs $H^{\pm}$ after electroweak symmetry breaking.  Typically, the $h^0$ has Standard Model-like couplings.  At the one-loop level, the Higgs boson mass is
\begin{eqnarray}
m_{h^0}^2 &\simeq& m_{Z^0}^2 \cos^2 2\beta 
\nonumber\\
&&+ \frac{3g^2m_t^4}{8\pi^2m_{W}^2}\left(
\log \frac{m_{\tilde{t}_1}m_{\tilde{t}_2}}{m_t^2} + a_t^2\left(1 - \frac{a_t^2}{12}\right)\right), \nonumber
\end{eqnarray}
where $a_t$ is the dimensionless trilinear coupling between the Higgs and top squarks
\begin{eqnarray}
a_t =  \frac{ A_t - \mu\cot \beta}{\sqrt{ \half (m^2_{\tilde{t}_1}+ m^2_{\tilde{t}_2}) }}.
\end{eqnarray}
For a moderate $a_t\lsim 1$ and top squarks lighter than 1 TeV, the Higgs mass is less than 120 GeV \cite{Amsler:2008zzb,Degrassi:2002fi}.  By taking $a_t$ to ``maximal mixing,'' where the contribution from the $A$-terms gives the largest contribution to the Higgs mass, the Higgs can be as heavy as 130 GeV while keeping the top squarks under 1 TeV.  Two-loop corrections can raise the Higgs mass by an additional $\lesssim$ 6 GeV \cite{Degrassi:2002fi}.  

To avoid fine tuning, the top squarks should not be significantly heavier than the Higgs.  Even with masses at 1 TeV, the Higgs potential is tuned at the few percent level.  If the top squarks are at 400 GeV, the fine tuning of the Higgs potential drops substantially; however, the upper limit on the Higgs mass falls to 120 GeV even with maximal top squark mixing \cite{MSSMFineTuning}.  This has motivated studies giving the Higgs quartic coupling additional contributions inside the supersymmetric Standard Model \cite{NewQuartics}, which usually leads to a less minimal Higgs sector such as in the  next-to-minimal supersymmetric Standard Model (NMSSM).

Alternate models of electroweak symmetry breaking that can have naturally light Higgs bosons are motivated by the indirect bounds coming from electroweak constraints and the desire to minimize fine tuning in the Higgs sector.  These models, which often have more elaborate Higgs potentials with additional scalar fields, allow light Higgs masses, while simultaneously evading the LEP direct bound.  Frequently, these less minimal models of electroweak symmetry breaking have approximate global symmetries and light pseudo-Goldstone bosons that can alter the phenomenology of the Higgs.  These light pseudo-Goldstone bosons can evade all existing limits because they couple very weakly to light flavor fermions.  

In this paper, we will focus on such non-minimal models of electroweak supersymmetry breaking.  We begin in Section II with a brief discussion of Higgs models that contain light pseudo-Goldstone bosons, focusing primarily on current experimental constraints.  In Section III, we propose a new search for Higgs bosons that cascade decay to pseudoscalars with masses below 10 GeV.  In particular, we find that the Higgs can be discovered in a subdominant decay mode where one pseudoscalar decays to muons and the other to taus.  We conclude with a discussion of the expected sensitivity to the Higgs production cross section at the Tevatron and LHC.  The proposed search would allow possible discovery of a cascade-decaying Higgs with the complete Tevatron data set or early data at the LHC.      

 \section{Light $a^0$ Modifications to Higgs Phenomenology}
 
In this section, we explore the couplings of a light pseudo-Goldstone boson, or ``axion,'' to the Standard Model.  We will describe how to analyze a general theory and we find constraints on the maximal width for a Higgs decaying into a light axion.  Finally, we will analyze how CLEO limits on the direct coupling between pseudo-Goldstones and Standard Model fermions set constraints on the Higgs width into axions.  For Type II two Higgs doublet models ({\em i.e.}, MSSM), the CLEO and LEP results place important limits on the light axion scenario.

The tension between the LEP limit on the Higgs mass and fine tuning can be reduced if the Higgs branching fractions are altered from those of the Standard Model \cite{Chang:2008cw}.  The LEP bound of $114$ GeV only applies when the Higgs decays dominantly to a $b \bar{b}$ or $\tau^+ \tau^-$ pair.  While adding new, invisible decay modes does not help because bounds for such processes are just as strong \cite{:2001xz}, other nonstandard decays remain open possibilities.  Consider the case where the Higgs decays dominantly to two new scalars $\phi$, which in turn decay to SM particles:
\begin{eqnarray}
h^0 \rightarrow  \phi\phi \rightarrow (X\bar{X})(X\bar{X}).
\end{eqnarray}
For an $h^0$ with SM-like production cross section, this process is excluded for $m_{h^0} < 110$ GeV and $X = b$ \cite{Abbiendi:2004ww,Abdallah:2004wy,Schael:2006cr}.  However, there is an 82 GeV model-independent bound from LEP \cite{Abbiendi:2002qp} and when $X = g, c, \tau$, there are no limits for Higgs masses above 86 GeV \cite{Abbiendi:2002in}.

The decay width of a 100 GeV Higgs into Standard Model particles is $\Gamma(h^0\rightarrow\text{ SM})\simeq$ 2.6 MeV; because the decay width is so small, the Standard Model decay mode is easily suppressed by the presence of new decay modes.  Any new light particle with $\OO(1)$ coupling to the Higgs will  swamp the decay modes into SM particles.  Many theories, such as little Higgs models and non-supersymmetric two Higgs doublet models, have light neutral states for the Higgs to decay into.  This phenomenon arises when there is an approximate symmetry of the Higgs potential that is explicitly broken by a small term in the potential.  There is a resulting light pseudo-Goldstone boson that couples significantly to the Higgs boson.  The Peccei-Quinn symmetry of a two Higgs doublet model is one such example.  In the MSSM, it is possible to have a light $A^0$, even with radiative corrections included \cite{Dermisek:2008dq}.   More often, there is an additional singlet, $S$, and an approximate symmetry that acts upon the Higgs boson doublets as $H_i \rightarrow e^{i\theta q_i} H_i$ with the singlet compensating by $S\rightarrow e^{i\theta q_s}S$.  During electroweak symmetry breaking, $S$ also acquires a vev, spontaneously breaking the symmetry; the phase of $S$ becomes a pseudo-Goldstone boson and has small interactions with the Standard Model when $\langle S\rangle \gg v$.
 
 For specificity, let us consider a two Higgs doublet model with an additional complex singlet.\footnote{A similar analysis was performed for a one Higgs doublet and a complex singlet in \cite{Barger:2008jx}.  A one Higgs doublet model with a light pseudoscalar typically does not have a large branching ratio of the Higgs into pseudoscalars.}  All three scalar fields acquire vacuum expectation values: $v_u= v \sin \beta$, $v_d=v \cos \beta$, and $\langle S \rangle$.  The interactions of the pseudo-Goldstone bosons can be described in the exponential basis
 \begin{eqnarray}
 \nonumber
H_u &=& 
\left(\begin{array}{c}
\omega^+ \sin\beta + h^+ \cos \beta\\
\frac{1 }{\sqrt{2}}(v\sin\beta + h_u)
\end{array}
\right) e^{ i\frac{a_u}{v\sin\beta}} \\
\nonumber
H_d &=&
\left(\begin{array}{c}
\frac{1 }{\sqrt{2}}(v\cos\beta + h_d) \\
- \omega^- \cos\beta + h^- \sin \beta
\end{array}
\right)
e^{i \frac{a_d}{v\cos\beta}} \\
S &=& \frac{1 }{\sqrt{2}} (\langle S\rangle +s^0)e^{i \frac{a_s}{\langle S\rangle}} .
 \end{eqnarray}

The pseudoscalar fields $a_u, a_d$ and $a_s$ get interactions either through derivative couplings from the kinetic terms or through explicit symmetry breaking.  One linear combination of the pseudoscalars,
\begin{equation}
\omega_{Z^0} = -a_u\sin\beta + a_d \cos\beta,  \nonumber
\end{equation}
becomes the longitudinal component of the $Z^0$.  The two other combinations are physical fluctuations that get mass through symmetry breaking effects in the Higgs potential.   Any terms in the Higgs potential proportional to $|H_u|^2$, $|H_d|^2$, or $|S|^2$ do not affect the mass or interactions of the pseudo-Goldstones and there are only a handful of possibilities for explicit symmetry breaking.  As an example, consider adding to the potential a sizeable coupling
 \begin{eqnarray}
V_1= \lambda_1  S^2 H_u^\dagger H_d^\dagger  +\hc.
 \end{eqnarray}
 This will give a weak-scale mass to the following linear combination of pseudo-Goldstones:
 \begin{eqnarray}
A^0=   \cos\theta_a( a_u \cos \beta + a_d \sin\beta)  -  a_s \sin\theta_ a.
 \end{eqnarray}
The singlet mixing component is given by
 \begin{eqnarray}
 \tan\theta_a = \frac{v}{\langle S \rangle} \sin 2\beta. 
\end{eqnarray}
The remaining linearly independent  pseudo-Goldstone will be massless until the final symmetry is broken.  This linear combination is
 \begin{eqnarray}
a^0 = \sin\theta_a (a_u \cos\beta + a_d \sin \beta)  + a_s \cos \theta_a
 \end{eqnarray}
and gets a mass through potentials such as
 \begin{eqnarray}
 \label{Eq: Symmetry Breaking}
 V_2 = \lambda_2 S^2 H_u H_d +\hc.
 \end{eqnarray}
There will be mixing between $a^0$ and $A^0$.  However, when $\lambda_1 \gg \lambda_2$, $A^0$ and $a^0$ are nearly mass eigenstates with residual mixing
 proportional to $m^2_{a^0}/m_{A^0}^2$.  
 
 It is worth noting that as $\langle S\rangle \ll v$, the light pseudoscalar becomes the Peccei-Quinn pseudoscalar and its couplings are independent of $\langle S\rangle$.
 This particular example has the same symmetry structure as the NMSSM near the R-symmetric limit  when $SH_u H_d$  $A$-term dominates the $S^3$ $A$-term.    We will couch our discussions in terms of the R-symmetric NMSSM for comparison with the literature, but other realizations of the symmetry breaking are just as applicable.  
 
  \begin{figure}[t] 
\centering
\includegraphics[width=3.5in]{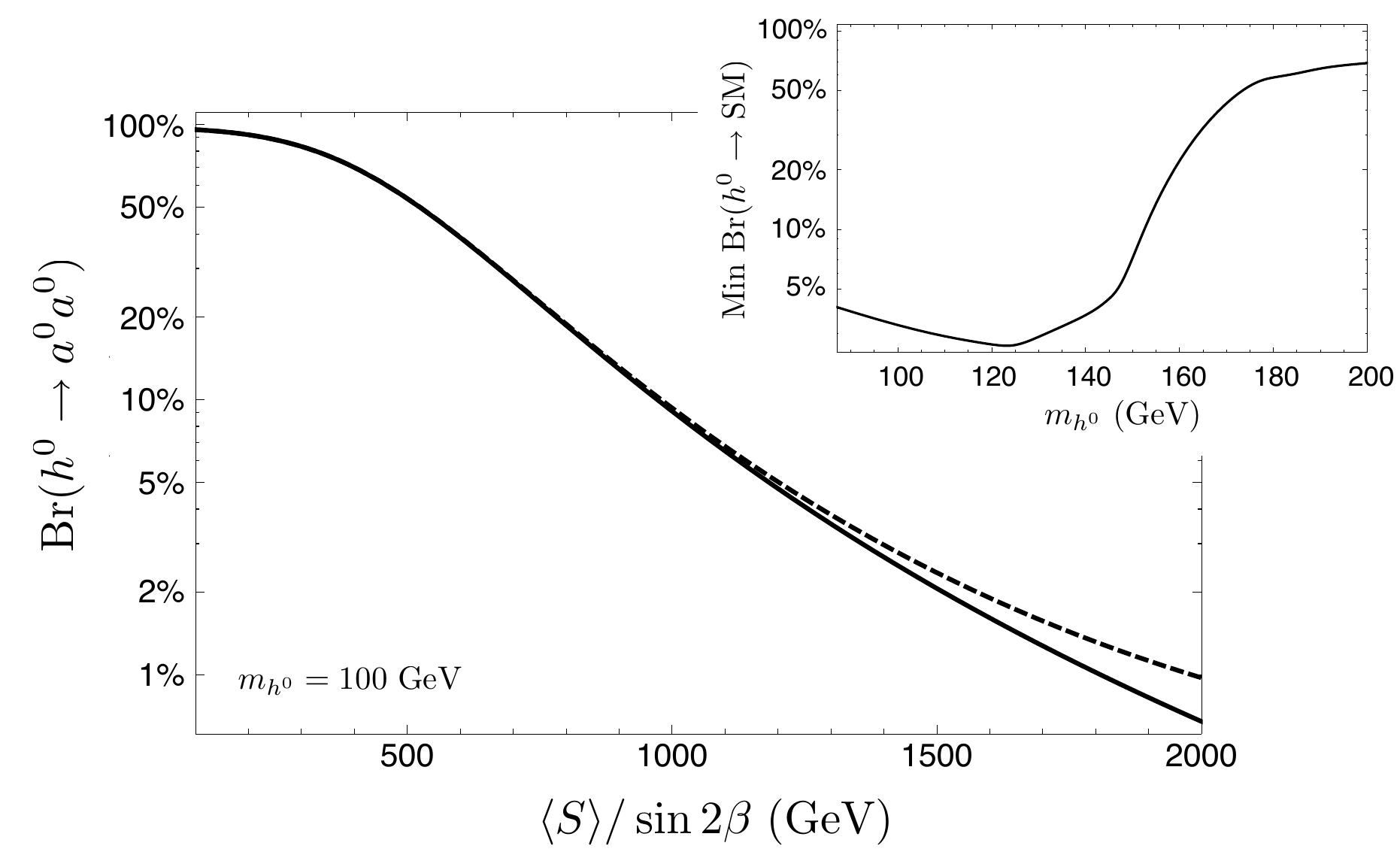}
 \caption{
 \label{Fig: Branching}
  The branching fraction of the Higgs into pseudoscalars as a function of $\langle S\rangle/\sin 2\beta$ for $m_{h^0}=100\GeV$ when $\tilde{d}_h=0$ and $1$ (solid and dashed lines, respectively).   The inset shows the minimum value of the branching rate into the Standard Model as a function of $m_{h^0}$.}
\end{figure}

To evade limits from LEP, the Higgs needs a significant branching rate into the pseudoscalar (Fig. 1), and one might worry that radiative corrections from the Higgs-pseudoscalar interaction might induce a large radiative correction to the pseudoscalar mass.  However, the interaction that leads to the Higgs decay into pseudoscalars can occur even if the axion is an exact Goldstone boson through the coupling
\begin{eqnarray}
\label{Eq: Decay Interaction}
\LL_{\text{int}} = \tilde{c}_h \frac{v}{\langle S \rangle^2}  h^0 \partial_\mu a^0 \partial^\mu a^0 - \tilde{d}_h \frac{ m_{a^0}^2}{v} h^0 a^0 a^0.
\end{eqnarray}
The first interaction preserves the $a^0 \rightarrow a^0 + \epsilon$ shift symmetry, where $\tilde{c}_h$ is an $\OO(1)$ constant and $v = 246$ GeV is the electroweak scale.  Because this coupling exists in the symmetry-preserving limit, $\tilde{c}_h$ can only depend on the vevs of the Higgs fields, the particular charges of the approximate $U(1)$ symmetry, and on the alignment of the physical Higgs boson relative to the Higgs vev direction.  
When the physical Higgs boson is in the direction of the Higgs vev, $h^0 = h_u \sin\beta + h_d \cos\beta$,  the example above gives
\begin{eqnarray}
\label{Eq: Higgs Axion Coupling}
\tilde{c}_h = \sin^2 \theta_a \frac{\langle S\rangle^2}{ v^2} = \frac{\sin^2 2\beta}{1+ \frac{v^2 \sin^2 2\beta}{\langle S\rangle^2}}  \simeq \frac{4}{\tan^2\beta} .
\end{eqnarray}
The second interaction breaks the shift symmetry and is proportional to $m_{a^0}^2$.  This term depends on the symmetry breaking that gives the axion a mass and is therefore model-dependent.  When the physical Higgs boson aligns with the Higgs vev, the symmetry breaking coupling simplifies to
\begin{eqnarray}
\tilde{d}_h = 1
\end{eqnarray}
for the potential in Eq. \ref{Eq: Symmetry Breaking}.   For a symmetry breaking potential
\begin{eqnarray}
V_{2'} = \lambda_{2'} S^4+\hc,
\end{eqnarray}
 $\tilde{d}_h$ would be small, arising from the residual mixing between $s^0$ and $h^0$.  
 
 It is possible to increase $\tilde{d}_h$ by having multiple terms in the potential contribute to $m_{a^0}^2$, with the pseudoscalar mass being less than either of the contributions.  For instance, with $V_2$ and $V_{2'}$
\begin{eqnarray}
\tilde{d}_h =  \frac{ 1}{1 + \frac{2 \lambda_{2'} \sin 2\beta}{ \lambda_2 \tan^2\theta_a}} \gsim  1 \quad \text{ if }  \lambda_{2'} \simeq -\frac{ \lambda_2 \tan^2\theta_a}{2 \sin 2\beta}.
\end{eqnarray}
   Of course, this is the technical definition of fine tuning and when $\tilde{d}_h \gsim 1$, $a^0$ has been fine tuned to be light.  We will discount this possibility in our discussion on the expected sizes of couplings in this class of theories.
 
The partial width of the Higgs into pseudoscalars from these interactions is
\begin{eqnarray}
\label{Eq: Partial Width}
\frac{\Gamma_{\! h^0\rightarrow a^0 a^0}}{m_{h^0}} = \frac{m^2_{h^0}}{16\pi}\!\!\left(\!\!\frac{  \tilde{c}_hv }{2\langle S \rangle^2} + \frac{ \tilde{d}_h m_{a^0}^2}{ v m^2_{h^0}}\!\! \right)^{\!2}\!\!\!\left(\!\!1+ \OO\Bigg(\!\frac{m_{a^0}^2}{m_{h^0}^2}\!\Bigg)\!\!\right).
\end{eqnarray}
\label{eq: partialwidth}
  The symmetry-preserving interaction dominates when
\begin{eqnarray}
\langle S\rangle  \le \left(\!\frac{ \tilde{c}_h}{2 \tilde{d}_h}\!\right)^\half\!\! \frac{m_{h^0} v }{ m_{a^0} } 
\lsim 1.5 \TeV.
\end{eqnarray}
When $\langle S \rangle$ is less than 1 TeV, the Higgs boson has an appreciable width into pseudoscalars (Fig. 1).  This is precisely the region we are interested in, so we will set $\tilde{d}_h=0$ for the rest of this discussion.

\begin{figure}[t] 
\centering
\includegraphics[width=3.5in]{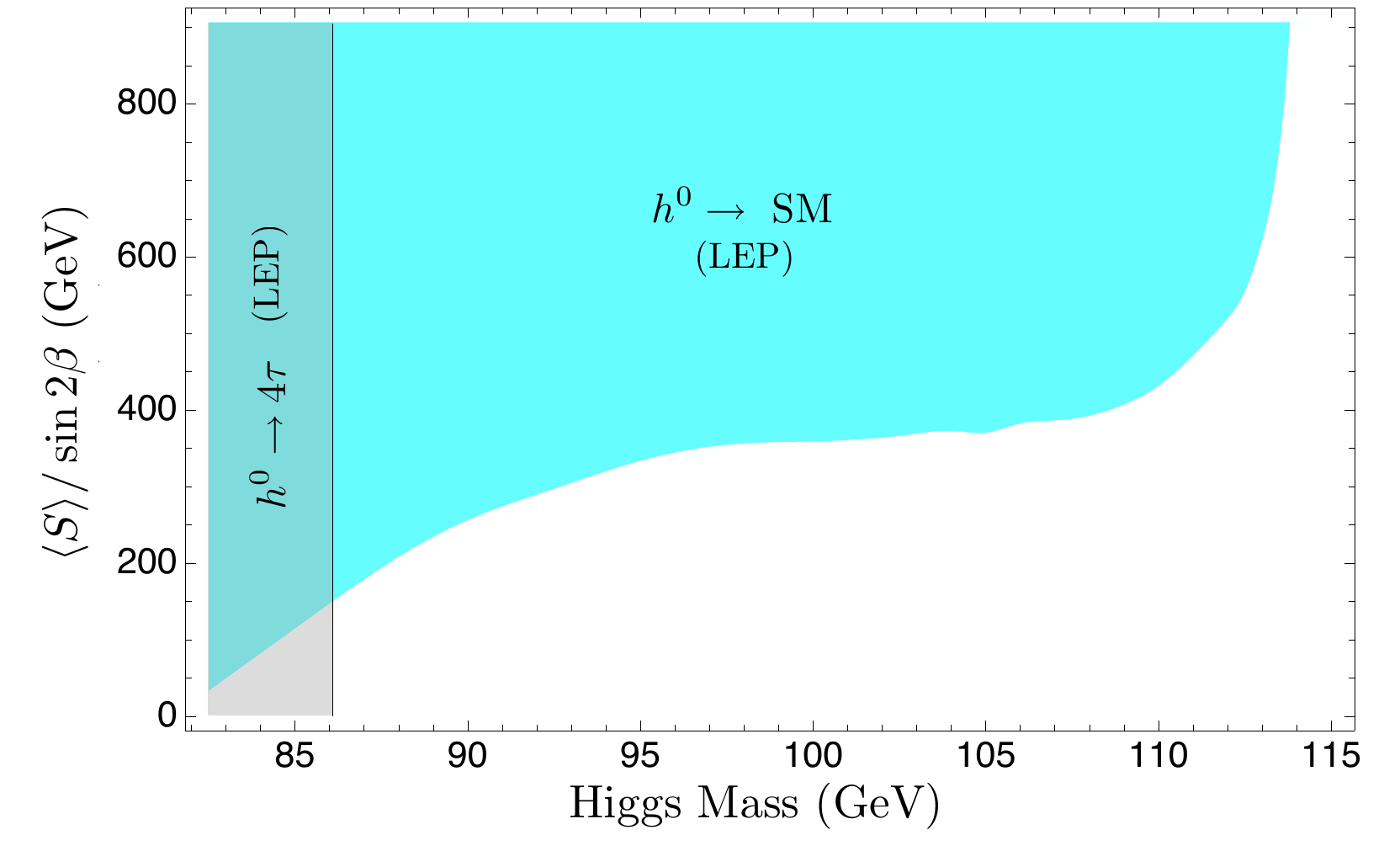}
 \caption{
 \label{Fig: HiggsLimit}
     Values of $\langle S \rangle/\sin2\beta$ (GeV) that have been excluded through LEP2's search for a Standard Model Higgs \cite{Barate:2003sz}.  The region below $m_{h^0} =86$ GeV is entirely excluded by the $h^0 \rightarrow 4 \tau$ search \cite{Abbiendi:2002in}.}
\end{figure}

Figure \ref{Fig: HiggsLimit} shows the values of $\langle S\rangle/ \sin 2\beta$  that are necessary to evade LEP2's search for a Standard Model Higgs \cite{Barate:2003sz}.   
For Higgs boson masses less than 100 GeV, the $a^0$ has to be fairly strongly coupled to the Higgs boson, requiring $\langle S \rangle$ to be small, which increases the size of the coupling to Standard Model fermions.  However, bounds from the recent CLEO results \cite{:2008hs} are strongest for large fermion couplings.  The CLEO bounds place a 90\% C.L. upper limit on Br($\Upsilon \rightarrow \gamma a^0$) Br($a^0 \rightarrow \tau^+ \tau^- $).  For $m_{a^0}$ between 3.5 GeV and 9 GeV, this limit ranges from $\sim 10^{-5} - 10^{-4}$ for the tau decays.  The branching fraction for radiative $\Upsilon$ decays is \cite{Hiller:2004ii}
\begin{equation}
\frac{\text{Br}(\Upsilon \rightarrow a^0 \gamma)}{\text{Br}(\Upsilon\rightarrow \mu^+ \mu^-)} = \frac{G_F m_{\Upsilon}^2}{4 \sqrt{2} \pi \alpha} g_d^2 \Bigg(1-\frac{m_{a^0}^2}{m_{\Upsilon}^2}\Bigg) F
\end{equation}
where $F$ is a QCD correction factor $\simeq 0.5$, $\text{Br}(\Upsilon(1s) \rightarrow \mu^+ \mu^-) = 2.5\%$, and $g_d$ is the axion coupling to down quarks.  For Type II two Higgs doublet models,\footnote{This case applies to the MSSM and its extensions.  For Type I two Higgs doublet models, where all Standard Model fermions only couple to one Higgs doublet, there is no asymmetry between up and down-type quarks in the coupling to axions and this typically results in a $\cot\beta$ suppression in the coupling to axions.}   the axion coupling to fermions is given by 
\begin{equation}
\LL_{\text{int}}=i g_{f} \frac{m_f}{v} \bar{f} \gamma_5 f a^0,
\end{equation}
where
\begin{eqnarray}
g_{f} = \sin \theta_a \begin{cases}
\cot\beta  & \mbox{      (up-type quarks)} \\ 
\tan\beta & \mbox{     (down-type quarks/leptons)}
 \end{cases}
\end{eqnarray}
 (see also  \cite{Dobrescu:2000yn,Dermisek:2006wr}).   The coupling to up-type quarks is suppressed by two powers of $\tan\beta$.  This means that, above the $b$-quark threshold, the axion will preferentially decay to $b$-quarks.  Below this threshold, it will preferentially decay to tau leptons, rather than charm quarks. 

\begin{figure}[b] 
   \centering
   \includegraphics[width=3.5in]{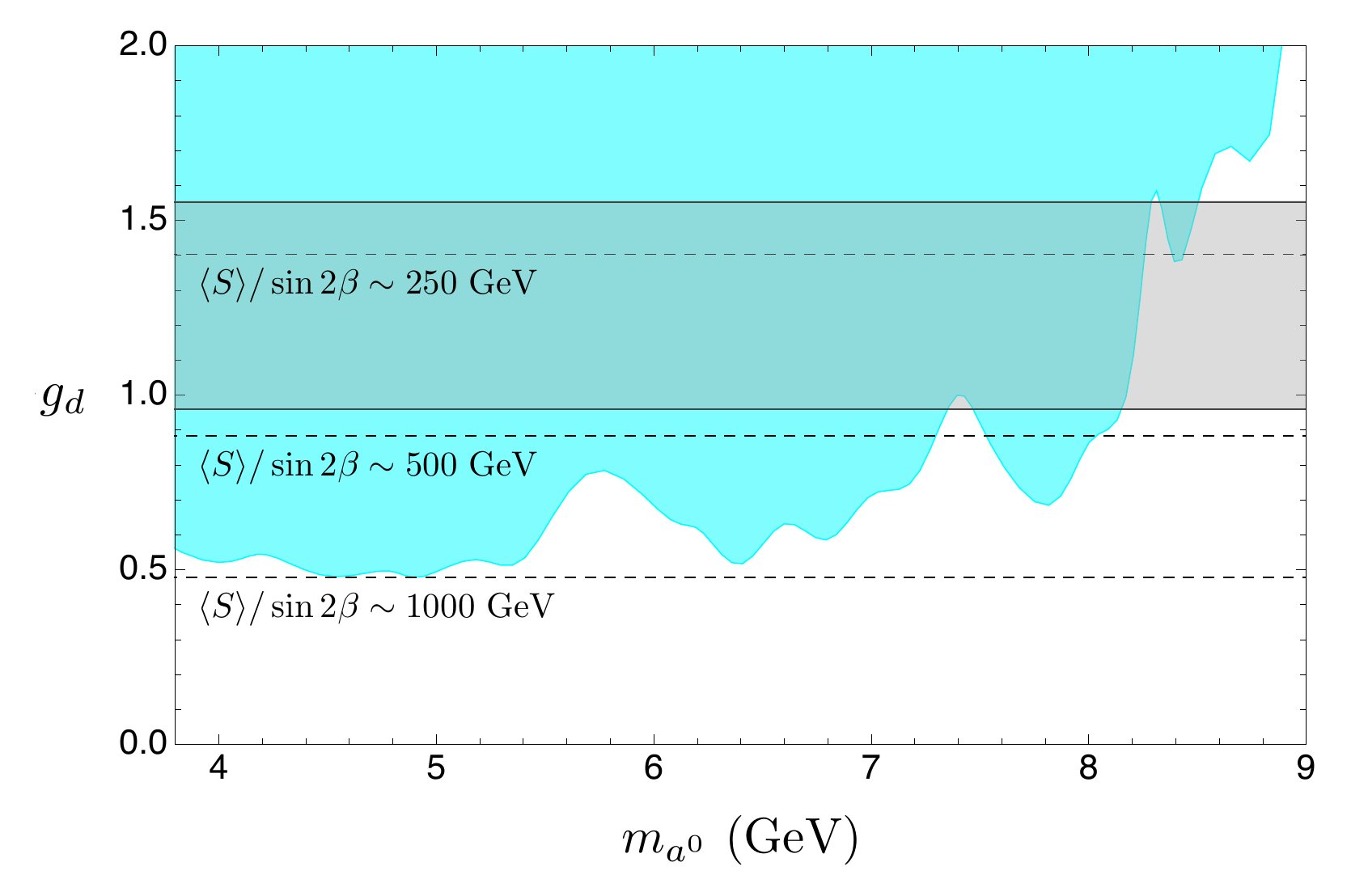} 
     \caption{ 
 Region of $m_{a^0} - g_d$ parameter space that has been excluded by CLEO to 90\% C.L  \cite{:2008hs}.  The dashed lines indicate values of $\langle S \rangle /\sin 2\beta$ for $\tan\beta = 2$.  The shaded region shows the minimum values of $g_d$ allowed by LEP for an 87-110 GeV Higgs.}
   \label{fig:example}
\end{figure}

The CLEO bound on the branching fraction of $\Upsilon$ sets a bound on $g_d$, which can be used to set limits on the allowed range of $g_d$ vs. $m_{a^0}$ (Fig.~\ref{fig:example}).  The CLEO results place the strongest constraints on small values of $\langle S \rangle/\sin2\beta$.  The value of the singlet vev is a measure of the fine tuning of the theory because it induces a mass for the Higgs bosons, $m^2_{\text{eff}} = \lambda \langle S \rangle^2$ in the scalar potential and with an $\OO(1)$ coupling, the singlet vev should be less than a few TeV to avoid large fine tuning \cite{NMSSMFineTuning}.    There is some tension, then, between keeping the coupling to fermions small and keeping the coupling to the Higgs boson sufficiently large to evade LEP limits without fine tuning the $a^0$ to be light.   

It is also possible for LEP to have directly produced the Higgs through $e^+ e^- \rightarrow Z^0\rightarrow h^0a^0$  \cite{Schael:2006cr}.  The LEP searches place bounds on the product of the squared $Z^{0} h^0 a^0$ coupling and the branching ratio of the Higgs into a Standard Model fermion $f$:
\begin{eqnarray}
\label{Eq: Z aH Limit}
\xi^2\simeq \frac{\sin^2\theta_a \sin^2 2\beta}{1+ \frac{1}{12}\frac{m_{h^0}^2}{m_b^2} \sin^4\theta_a} \!\!\Br(h^0\rightarrow f\bar{f})_{\text{SM}} \le \frac{\sqrt{3} m_b}{m_{h^0}},
\end{eqnarray}
where $\Br(h^0\rightarrow f\bar{f})_{\text{SM}}$ is the Standard Model's branching ratio to fermion pairs.
 There were searches for the $(b\bar{b})(\tau^+\tau^-)$ final state at LEP, but there were no limits for $75\GeV \le m_{h^0}\le 125 \GeV$.  For $125\GeV \le m_{h^0}\le 165\GeV$, the limits were $\xi^2 \lsim \OO(0.4)$ which is automatically satisfied in these models.  There were additional searches for the $(\tau^+ \tau^-)(\tau^+ \tau^-)$ final state, but these constraints are even weaker because the Higgs branching fraction into taus is a factor of ten smaller than that into bottoms.  LEP did search for  $e^+ e^- \rightarrow Z^0\rightarrow h^0a^0\rightarrow (a^0 a^0)a^0$, but the search for the $6\tau$ final state was only performed at LEP1 and was thus not sensitive to Higgs masses above 75 GeV.
 
\section{$h^0\rightarrow a^0a^0$ at Hadron Colliders}

We will now discuss how the Higgs can be discovered if it decays into a light pseudoscalar $a^0$, when $2m_{\tau} \lesssim m_{a^0} \lesssim 2 m_b$.  In this range, the $a^0$ decays predominantly into taus and the signature of the Higgs is the appearance of 4$\tau$ events.  All existing searches for this decay channel have focused on the scenario where two or more taus decay leptonically \cite{Graham:2006tr, :2008uu,Forshaw:2007ra}.   Currently, the ATLAS collaboration is exploring the $4 \mu 8 \nu$ channel and CMS is analyzing $(\mu^{\pm} \tau^{\mp}_{h})(\mu^{\pm}\tau^{\mp}_h)$ \cite{:2008uu}.  There are specific challenges to the $4 \tau$ decay channel, however.  The branching fraction of the taus to leptons is only 33\% and the $p_T$ spectrum of the events is soft because the visible lepton carries less than half the momentum of the tau.  Additionally, it is challenging to reconstruct the Higgs and pseudoscalar masses from the final decay products.

\subsection{Signal}

The primary innovation of the search proposed in this paper is to use the subdominant decay of the $a^0$ into two muons, which exists because the $a^0$ couples to the Standard Model by mixing through the CP-odd Higgs.  The relative branching ratio for the $a^0$ into muons versus taus is
\begin{eqnarray}
\frac{\Gamma(a^0\rightarrow \mu^+\mu^-)}{\Gamma(a^0 \rightarrow \tau^+\tau^-)} = \frac{ m_\mu^2}{m_\tau^2 \sqrt{1- (2 m_\tau/m_{a^0})^2}}.
\end{eqnarray}
The cross section of $h^0 \rightarrow 2\mu 2\tau$ depends upon the following product of branching ratios:
\begin{eqnarray}
\epsilon_{\mu\tau}=2  \Br(a^0 \rightarrow \mu^+ \mu^-) \Br(a^0 \rightarrow \tau^+ \tau^-).
\label{eq: factor}
\end{eqnarray}
For $\tan \beta\gsim 4$, a 7 GeV pseudoscalar has a 0.4\% branching ratio into muons and 98\% ratio into taus.   As $m_{a^0}$ goes from the bottom threshold to the tau threshold, $\epsilon_{\mu\tau}$ varies from 0.8\% to 1.5\%.  The remaining events go into hadrons and are divided between the charm and glue-glue decay channels.  For $\tan\beta=2$ the branching ratio to charms becomes 15\% and the branching ratio to taus and muons is reduced to 83\% and 0.3\%, respectively, causing $\epsilon_{\mu\tau}$ to fall to 0.2\% for a 7 GeV $a^0$ and 0.5\% for an $a^0$ just above the tau threshold.  The events that go into hadrons do not typically have significant missing energy and do not pass the missing energy cuts. 
 \begin{table}
\begin{center}
\begin{tabular}{|c||c|c|}
\hline
		&\multicolumn{2}{c|}{Signal Efficiency}\\	\hline
Selection Criteria		& Relative 	& Cumulative			\\ 	\hline
Pre-Selection Criteria 	& 26\% 		& 26\%				\\
Jet veto				& 99\%		& 26\%	\\
Muon iso \& tracking		& $\sim 50\%$	& 13\%			\\					
$M_{\mu\mu} < 10$ GeV & 98\%		& 13\%				\\ \hline
 $p_T^{\mu\mu}> 40$ GeV& 76\%		& 9.8\%				\\
$\MET>30$ GeV		& 29\%		& 2.8\%				\\
$\Delta\phi(\mu,\MET)>140^{\circ}$&73\%& 2.1\% 				\\ 	
$\Delta R(\mu,\mu)>$0.26	& 63\%		& 1.8\%				\\
\hline
\end{tabular}
\caption{\label{Tab: Cuts} Relative and cumulative signal efficiencies due to the specified selection criteria.  The signal point is a 100 GeV Higgs decaying to a 7 GeV $a^0$ at the LHC.  The pre-selection criteria include finding a pair of oppositely-signed muons, each with $|\eta| <2$ and $p_T > 10$ GeV.}
\label{Fig: cutstable}
\end{center}
\end{table}

Due to the pseudoscalar's small branching fraction into muons, this decay channel has not been explored.  However, the small branching fraction into muons need not be a deterrent.  The main contribution to the cross section for light ($\sim 100$ GeV) SM-like Higgses comes from gluon-gluon fusion
and can be as high as $2 \text{ pb}$ at the Tevatron or $50 \text{ pb}$ at the LHC  \cite{Amsler:2008zzb}.  As a result, it is still possible to get $300$ events with 20 fb$^{-1}$ at the Tevatron (combined $\DO$ and CDF) and $250$ events per experiment at the LHC (at $\sqrt{s}=14\TeV$) with  500 pb$^{-1}$ luminosity, despite the small branching fraction to muons.    We ultimately find $\OO(2\%)$ cumulative efficiency for the signal (Table I), resulting in 95\% exclusion limits in certain mass windows at the Tevatron.  At the LHC, there is the possibility for discovery within the first year of running.  

\begin{figure}[t] 
\centering
\includegraphics[width=3.5in]{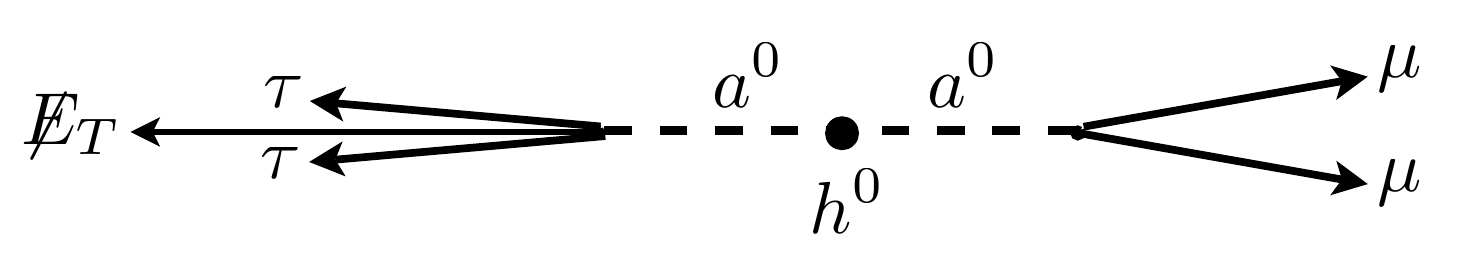} 
\caption{Schematic of Higgs decay chain.  The muons and taus will be highly boosted and nearly collinear.  It is likely that the taus will be reconstructed as one jet.  Most of the $\MET$ in the event will be in the direction of this jet.}
\label{fig:diagram}
\end{figure}

When the Higgs boson decays to two light CP-odd scalars $a^0$, the pseudoscalars are highly boosted and back-to-back in the center-of-mass frame (Fig.~\ref{fig:diagram}).  We consider the case where there is a nearly-collinear pair of oppositely-signed muons on one side of the event and a nearly-collinear pair of taus on the other, which we refer to as a ditau (di$\tau$).  Each tau has a $66\%$ hadronic branching fraction; consequently, there is a $44\%$ probability that both taus will decay into pions and neutrinos, which the detector will see as jets and missing energy.  Even if the taus do not both decay hadronically, there is still missing energy, as well as a jet and a lepton, except when both taus decay to muons, which occurs $\sim 3\%$ of the time.  The signal of interest is
\begin{equation}
p p\rightarrow  \mu^+ \mu^- + \text{di}\tau +\MET, \nonumber
\end{equation}
where the missing energy comes from the boosted neutrinos and points in the direction of the ditau.  Because the taus are nearly collinear, the ditaus are often not resolved, leading to a single jet-like object.

Signal events for a 7 GeV pseudoscalar decaying into $2\mu 2\tau$ ($\epsilon_{\mu \tau} =0.8\%$) were generated, showered, and hadronized using {\tt PYTHIA 6.4} \cite{Sjostrand:2006za}.\footnote{{\tt PYTHIA} does not keep spin correlations in decays.  This approximation does not affect the signal considered here because the taus are highly boosted in the direction of $a^0$ and any kinematic dependence on spin is negligible.  As verification of this, {\tt TAUOLA} \cite{Jezabek:1991qp} was used to generate the full spin correlated decays.}   
Unlike at LEP, the overall magnitude of the Standard Model Higgs production cross section is sensitive to physics beyond the Standard Model and it is possible to increase the cross section by an order of magnitude by adding new colored particles that couple to electroweak symmetry breaking.  In this study, the NNLO Standard Model production cross section was used as the benchmark value \cite{Amsler:2008zzb}.  

{\tt PGS} \cite{PGS} was used as the detector simulator.  Because the muons are adjacent, standard isolation cannot be used.  The muon isolation criteria must be modified to remove the adjacent muon's track and energy before estimating the amount of hadronic activity nearby.  As a result, we did not require standard muon isolation in this study and instead reduced the overall efficiency by a factor of 50\% to approximate the loss of signal events from modified isolation.

\subsection{Backgrounds}

There are several backgrounds to this search:  Drell-Yan muons recoiling against jets, electroweak processes, and leptons from hadronic resonances.  The Drell-Yan background is the most important.  The missing energy that results from the tau decays is a critical feature in discriminating the signal from the background.  In addition, the fact that the missing energy is in the opposite direction as the muons reduces the background from hadronic semileptonic decays. 

\begin{table}
\begin{center}
\begin{tabular}{|c||l|l|}
\hline
fb/GeV&  \multicolumn{1}{c|}{TeV}& \multicolumn{1}{c|}{LHC}\\
\hline\hline
DY+$j$&\;\quad 0.15&\,\quad0.24\\
$W^+W^-$&\;\quad0.03&\,\quad0.08\\
$t\bar{t}$&\;\quad 0.02& \,\quad0.14\\
$b\bar{b}$&$ \lsim 0.001$& \,$\sim 0.03$\\
$\Upsilon+j$& \;\quad$ 0.001$&\;\quad0.002\\
$\mu\mu$+$\tau\tau$& $\ll0.001$&$\lsim 0.001$\\
$J/\psi+j$& $\ll 0.001$& $\ll 0.001$\\
\hline\hline
Total& \;\quad 0.20&\;\quad 0.49\\
\hline
\end{tabular}
\caption{\label{Tab: Backgrounds}  Continuum backgrounds for low invariant mass muons pairs with missing energy ($d\sigma/dM_{\mu \mu}$) for the $h^0\rightarrow a^0a^0\rightarrow (\mu^+\mu^-)(\tau\tau)$ search at the Tevatron and LHC in units of \mbox{fb/GeV}.  The backgrounds are given for $p_T^{\mu \mu}, \MET$, and $\Delta R$ cuts optimized for a 100 GeV Higgs.  
}
\end{center}
\end{table}

The primary background arises from Drell-Yan muons recoiling against a jet.  The missing energy is either due to mismeasurement of the jet's energy or to neutrinos from heavy flavor semi-leptonic decays in the jet.  In the former instance, the analysis is sensitive to how {\tt PGS} fluctuates jet energies.  While {\tt PGS} does not parameterize the jet energy mismeasurement tail  correctly, the background only needs an $\OO(30\%)$ fluctuation in the energy, which is within the Gaussian response of the detector.  
The Drell-Yan background was generated using {\tt MadGraph}/{\tt MadEvent, v.4.4.16}\footnote{This version of {\tt MadEvent} does not apply the xqcut to leptons.  We thank J. Alwall for altering matrix element-parton shower matching for this study.} \cite{Alwall:2007st} and was matched up to $3j$ using an MLM matching scheme.  It was then showered and hadronized with {\tt PYTHIA}.  Again, the standard muon isolation criteria could not be applied and we used the same 50\% efficiency factor that was used for the signal.

All events are required to have a pair of oppositely-signed muons within $|\eta| < 2$.  Each muon must have a $p_T$ of at least 10 GeV.  A jet veto is placed on all jets, except the two hardest.  The veto is $15$ and $50$ GeV for the Tevatron and LHC, respectively.  Lastly, it is required that the hardest muon and missing energy are separated by $\Delta \phi \geq 140^\circ$.  Table I shows the relative and cumulative cut efficiencies for the signal.

There are three higher-level cuts that further distinguish the signal from the background.   These cuts are optimized as a function of the Higgs mass to maximize the significance of the signal.  The first is a cut on the sum $p_T$ of the muons ($p_T^{\mu \mu}$), and is approximately
\begin{equation}
p_T^{\mu\mu} \gtrsim 0.4 m_{h^0}.
\end{equation}
The second is a missing energy cut.  There is a moderate amount of missing energy in the signal events coming from the tau decays and this proves to be a very important discriminant from the Standard Model background.  The missing energy cut is 
\begin{equation}
\MET \gtrsim (0.2-0.25) \times m_{h^0}.
\end{equation}
For the LHC, the $\MET$ requirement is always held above 30 GeV.
The last is a $\Delta R$ cut on the muon pair, which depends on both the Higgs and pseudoscalar masses
\begin{equation}
\Delta R(\mu, \mu) \gtrsim \frac{4 m_{a^0}}{m_{h^0}}.
\end{equation}
These cuts depend on the kinematics of the decays and the geometry of the events is similar at both the Tevatron and LHC.
\begin{figure}[b] 
   \centering
   \includegraphics[width=3.5in]{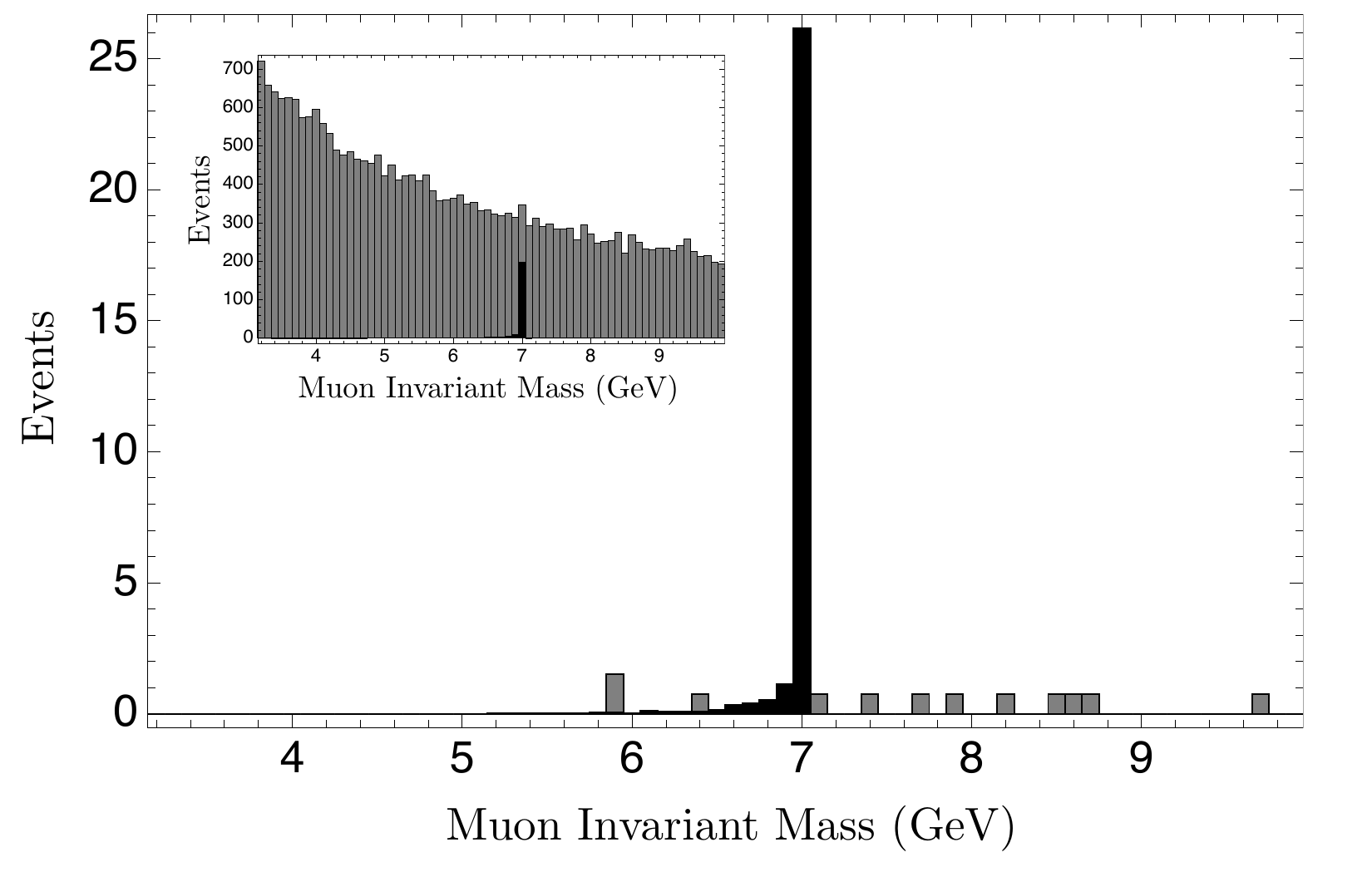} 
   \caption{Muon invariant mass for 5 fb$^{-1}$ at the LHC before (inset) and after the $p_T^{\mu\mu}$, $\MET$, and $\Delta R$ cuts.  The signal, a 100 GeV Higgs decaying to a pair of 7 GeV pseudoscalars, is shown in black and the Drell-Yan background is shown in gray.}
   \label{fig: invmass}
\end{figure}

Figure~\ref{fig: invmass} shows the invariant mass spectrum for the two oppositely-signed muons.  The inset shows the signal (black) and background (gray) before the $p_T^{\mu \mu}$, $\MET$, and $\Delta R$ cuts.  After these cuts are placed, the Drell-Yan background is mostly eliminated.  The muon invariant mass reconstructs the mass of the pseudoscalar.  

We used {\tt PGS} to model the muon invariant mass resolution and used an $m_{a^0}\pm 80 \MeV$ to exclude continuum backgrounds.  The Drell-Yan background ($d\sigma/dM_{\mu\mu}$) is $\OO(0.15 \text{ fb/GeV})$ at the Tevatron and $\OO(0.24 \text{ fb/GeV})$ at the LHC.\footnote{The background cross sections we quote in this section are for $p_T^{\mu \mu}, \MET,$ and $\Delta R$ cuts optimized for a 100 GeV Higgs.}

\begin{figure}[t] 
   \centering
   \includegraphics[width=3.5in]{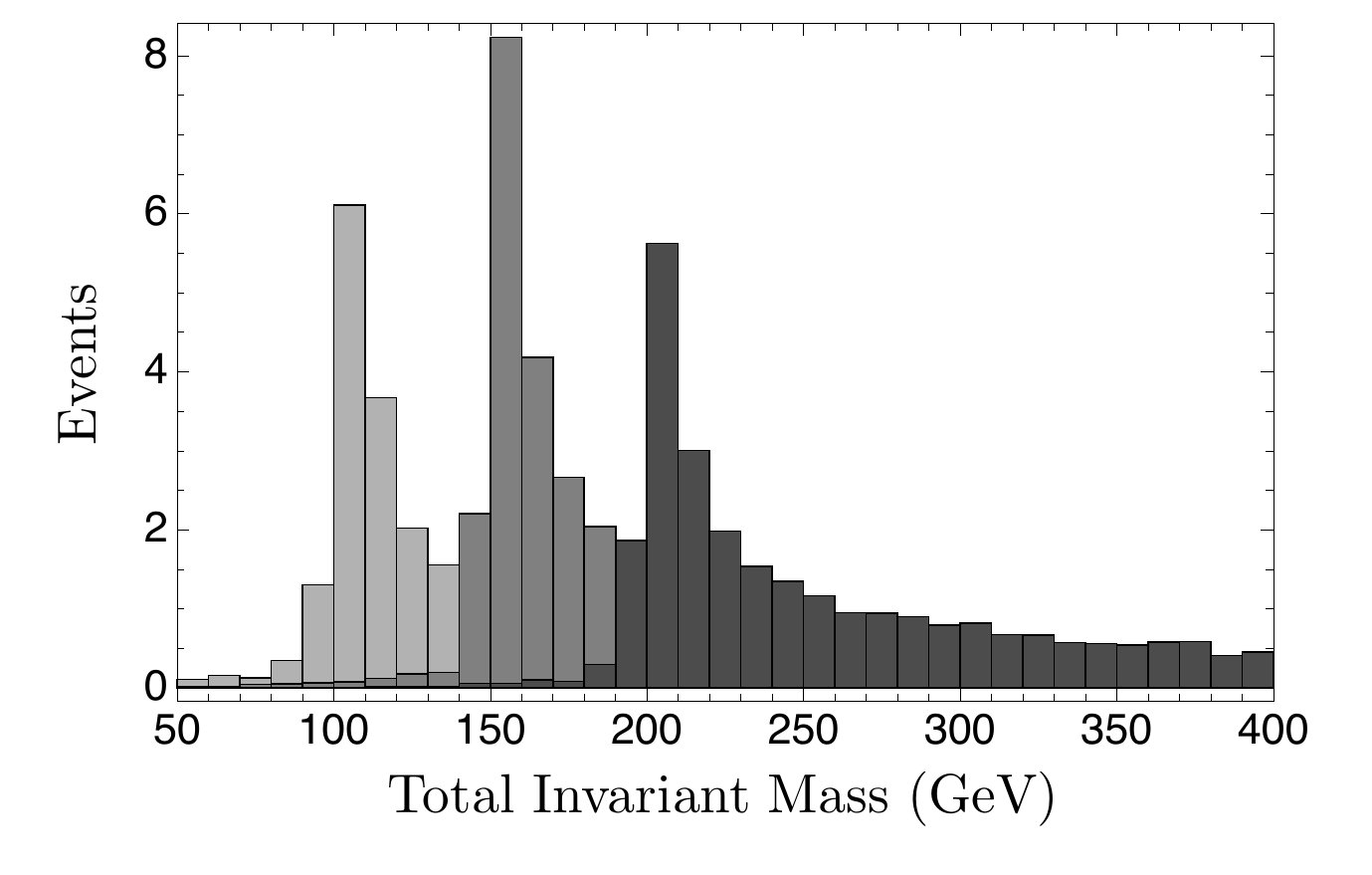} 
   \caption{Total invariant mass for signal with $m_{h^0} = 100, 150, 200$ GeV for 5 fb$^{-1}$ at the LHC after $p_T^{\mu\mu}$, $\MET$, and $\Delta R$ cuts.  The $\MET$ is projected in the direction of the hardest jet.}
   \label{fig: totinvmass}
\end{figure}

The other important kinematic handle in this analysis is the total invariant mass of the event, which reconstructs the mass of the s-channel Higgs boson.  The total invariant mass of the signal is shown in Fig. \ref{fig: totinvmass} after all cuts have been applied.  The width of the peak is narrowed if the missing transverse energy is projected in the direction of the jet.  We expect that this should be the direction of the missing energy because there will be boosted neutrinos from the hadronic tau decays.

In addition to Drell-Yan production, there are several electroweak production mechanisms for muon pairs and $\MET$.  The most important one is  $W^+ W^-$ production.  When the vector bosons are in a spin-0 configuration and decay leptonically, the muons are nearly collinear and antiparallel to the neutrinos.    When the $W^-$ decays to $\mu^-$, the lepton momentum and spin are in the same direction as the gauge boson.  The antineutrino, however, is antialigned with the $W^-$, and thus its momentum is antiparallel to that of the muon.  The situation is similar for the $W^+$ decay, except that the directions of the muon and neutrino are reversed.  The $\mu^+ \mu^- \nu \nu$ background was generated with {\tt MadGraph} and was found to be $\OO(0.03 \text{ fb/GeV})$ at the Tevatron and $\OO(0.08 \text{ fb/GeV})$ at the LHC.

Top quark production is another important electroweak background.  Using {\tt Madgraph} to generate $\mu^+ \mu^- \nu \nu b\bar{b}$, we estimate that this background is $\OO(0.02 \text{ fb/GeV})$ at the Tevatron and $\OO(0.14 \text{ fb/GeV})$ at the LHC.  The top background becomes nearly comparable to the Drell-Yan background for larger Higgs masses due to the weaker $p_T^{\mu \mu}, \MET,$ and $\Delta R$ cuts.

Electroweak production of $\mu \mu \tau \tau$ has a production cross section on the order of several attobarns when requiring low invariant-mass, high $p_T$ muons.  Consequently, it is subdominant to the $W^+W^-$ and $t \bar{t}$ backgrounds, with $\OO(8\times10^{-5} \text{ fb/GeV})$ at the Tevatron and $\OO(2 \times 10^{-4} \text{ fb/GeV})$ at the LHC.
\begin{figure*}[t] 
   \centering
         \includegraphics[width=3.25in]{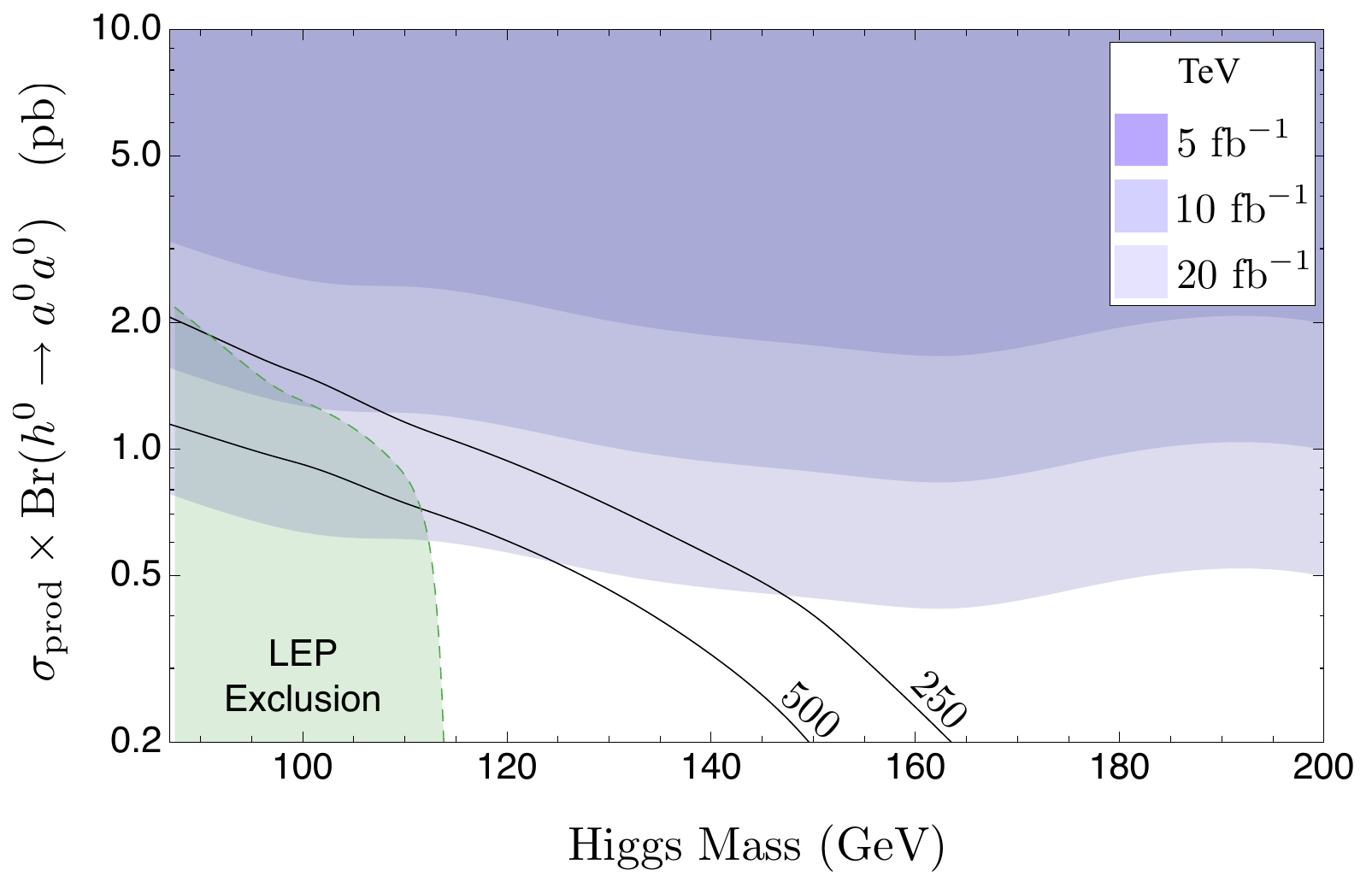} 
   \includegraphics[width=3.25in]{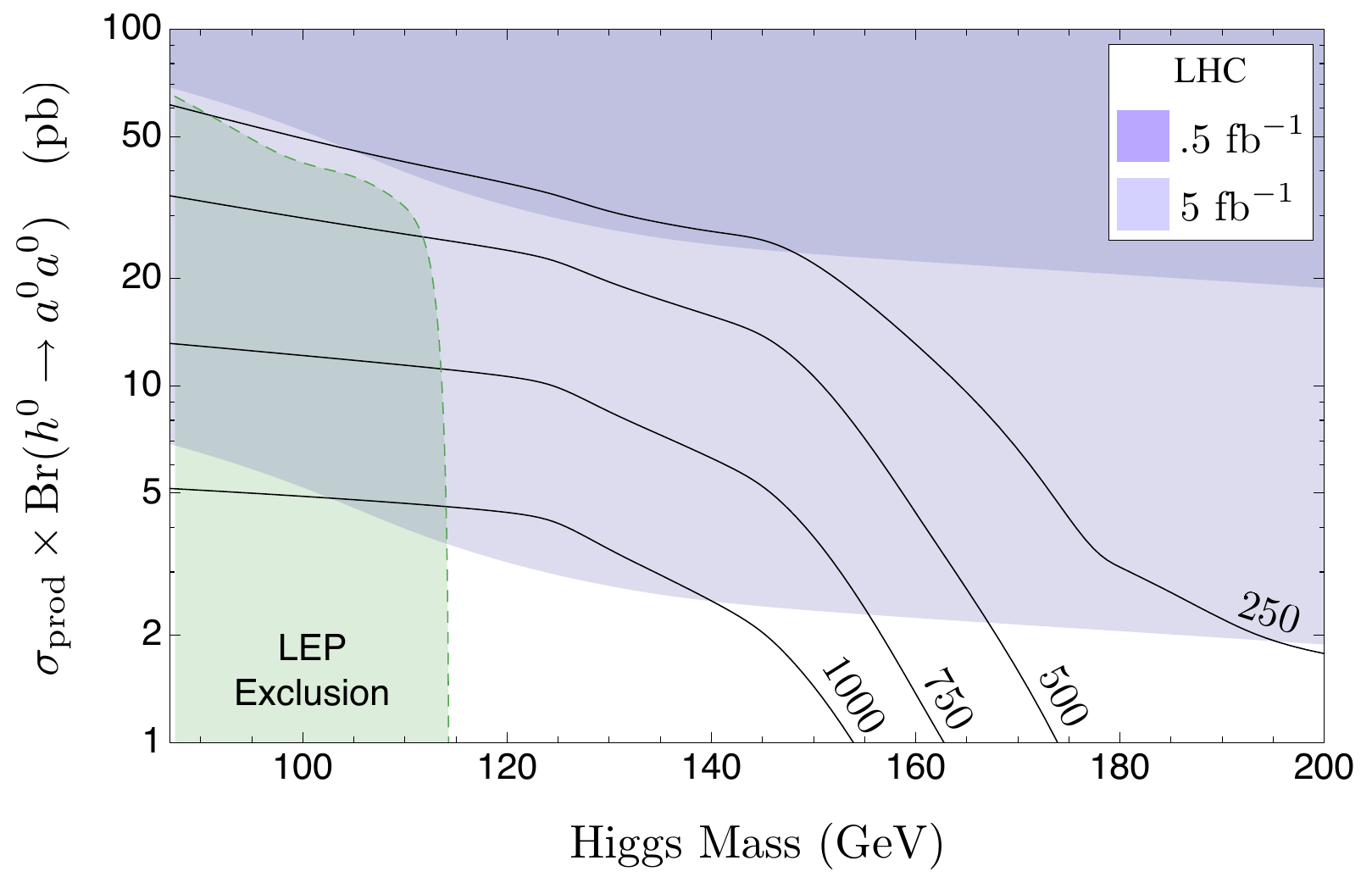} 
   \caption{Expected sensitivity to the Higgs production cross section at the Tevatron 
   (left) and LHC (right) for $m_{a^0} = 7$ GeV.  The contour lines indicate the cross sections for several values of $\langle S \rangle /\sin2\beta$ (in GeV), which alters Br($h^0\rightarrow a^0 a^0$).  The Standard Model Higgs decay width and (NNLO) gluon fusion production cross sections were obtained from \cite{Amsler:2008zzb}.  An $\epsilon_{\mu\tau}$ of 0.8\% was used for the branching ratio of $a_0a_0\rightarrow 2\mu 2\tau$.  The region beneath the dashed line has been excluded by LEP.}
   \label{fig: excplot}
\end{figure*}

There are several other important backgrounds that arise from low-lying hadronic spectroscopy that cannot be computed reliably with existing Monte Carlo generators.  These backgrounds come from  (i) semi-leptonic decays ($b \rightarrow c \mu \nu$), (ii) heavy-flavor quarkonia, and (iii) leptonic decays of light mesons.

Double semileptonic decays ({\it e.g.} $b\rightarrow c\rightarrow s/d$) typically give rise to soft leptons in jets but can occasionally fluctuate to give hard isolated leptons.  It is challenging to estimate this background contribution because the events are rare and we are statistics-limited.  However, we have attempted to estimate the relative magnitude using {\tt PYTHIA}.  It was found that the total cross section for $b\bar{b}$ jets to produce two muons is $\OO(80 \mbox{ } \mu\text{b})$ at the LHC.  Using a power law extrapolation from low $p_T^{\mu\mu}$, it was estimated that a $p_T^{\mu\mu}$ cut of 40 GeV reduces the cross section to $\OO(10 \text{ pb})$.  A missing energy cut of 30 GeV is 0.6\% efficient and requiring $\Delta \phi(\MET,\mu) > 140^\circ$ reduces the cross section by an additional order of magnitude to $\OO(5 \text{ fb})$.  Placing a $\Delta R$ cut on the muons and assuming that the muon isolation is 10\% efficient, the cross section becomes $\OO(0.3 \text{ fb})$.  The invariant mass of the muons in these events is distributed over $\OO(10 \text{ GeV})$, so the final background is approximately $\OO(0.03 \text{ fb/GeV})$.  This is likely an overestimate because we have assumed that the cuts are uncorrelated.  We found that at the Tevatron, semileptonic decays do not produce enough high $p_T$ muon pairs antialigned with the $\MET$ to be an important background. 

$\Upsilon$s can decay into muon pairs, but their invariant mass is above the range we are interested in.  $\Upsilon$s can also decay into taus that can subsequently decay into muons with a branching fraction of 3\%.  The invariant mass for these muon pairs will be in the region of interest. There are, however, two factors that mitigate this background.  The first is that the muons will be soft unless the $\Upsilon$ has very high $p_T^{\Upsilon} \sim \OO(60 \GeV)$.   The $p_T$ spectrum of $\Upsilon$s falls off rapidly.  At the Tevatron, the differential cross section is $\OO(250 \text{ fb})$ at $p_T^\Upsilon \sim 20 \GeV$ \cite{Acosta:2001gv}.   A na\"ive extrapolation to 60 GeV would place this background at $\OO(2\text{ fb})$.  Additionally, the missing energy in these events points in the direction of the muon pairs rather than towards the recoiling jet and a cut on the angle between the missing energy and the muon direction should reduce this background by another order of magnitude.  Accounting for this reduction, as well as the 3\% branching fraction of the taus to muons, we find that the cross section is $\OO(6 \times 10^{-4} \text{ fb/GeV})$ at the Tevatron.  We do not expect this background to dominate Drell-Yan at the LHC, either; using NNLO predictions for the $p_T$ distribution of $\Upsilon$s at the LHC \cite{Artoisenet:2008fc}, we estimate that this background will be $\OO(2 \times 10^{-3} \text{ fb/GeV})$. 

At the charm threshold,  the $J/\psi$ and $\psi(2S)$ become important because the tails of distributions arising from mismeasurement could spill over into higher invariant mass bins.  
Again, the cross section for the decays of these particles drops sharply as a function of $p_T^{\mu\mu}$ and with $p_T^{J/\psi}\ge 40\GeV$, the cross section is $\OO(100 \text{ fb})$ \cite{Abe:1997yz}.  Because this peak is below the invariant mass of interest, only the tail of the $M_{\mu\mu}$ distribution is a background.   The dominant contribution comes either from the Lorentzian tail of the decay width or from the non-Gaussian mismeasurement tail.   The Lorentzian tail suppresses the $J/\psi$ contamination  by $\OO(10^{-9})$ for  $m_{a^0}$ between 3.6 and 9 GeV.  The Gaussian tail of $J/\psi$ mismeasurement goes out at least 5$\sigma$, meaning that the contamination should be down by $\OO(10^{-6})$.   This gives a background cross section smaller than $\OO(10^{-5} \text{ fb/GeV})$ at the Tevatron and  at the LHC.  The contributions of the $\psi(2S)$ are subdominant to that of the $J/\psi$ \cite{Abe:1997jz}.  

Resonances beneath the $J/\psi$ are not a problem because they are far enough away from the invariant mass window we are interested in.
Peaks from  fake muons may arise from $B\rightarrow K\pi$  or similar decays where the kaons and pions punch through to the muon chamber.  These events are typically accompanied by significant hadronic activity and tight muon isolation requirements (after removing the adjacent muon) will reduce these backgrounds of fake muons \cite{Andy}.  Secondly, the $\MET$ from in-flight decays is in the direction of the muons, but in the signal, it is back-to-back with the muons.  Placing a cut on the relative angle between the muons and the $\MET$ is effective at eliminating these difficult backgrounds.

\subsection{Expected Sensitivity}
\label{Sec:Sensitivity}

Figure~\ref{fig: excplot} shows the expected 95\% exclusion plot at the Tevatron and LHC.  The contours indicate the cross sections for values of $\langle S \rangle/\sin2\beta$; this ratio affects the partial width of the Higgs into the pseudoscalars (Eq. 14).  The total projected luminosity for the combined data sets at CDF and $\DO$ is 20 fb$^{-1}$; currently, each experiment has $\sim 5$ fb$^{-1}$.  With 10 fb$^{-1}$ luminosity, the Tevatron will start probing the interesting regime where $\langle S\rangle/\sin2\beta= 250$ GeV.  Once the benchmark luminosity is reached, the Tevatron will have sensitivity up to $\langle S \rangle/\sin2\beta=500$ GeV.

With early data, the LHC has sensitivity to regions corresponding to $\langle S \rangle/\sin2\beta \lesssim 250$ GeV.  The sensitivity is weaker for Higgs masses below 100 GeV because the backgrounds worsen due to a smaller $p_T^{\mu\mu}$ cut.  However, combined analyses by CDF and $\DO$ should be able to probe this region down to a $\OO(1\text{ pb})$.  By the time the LHC reaches a luminosity of 5 fb$^{-1}$, it will be sensitive to the most relevant region of parameter space, with $\langle S \rangle/\sin2\beta \lesssim 1\TeV$. 

The sensitivity curves depend on the product of the pseudoscalar branching ratios into muons and taus, $\epsilon_{\mu \tau}$.  For Fig.~\ref{fig: excplot}, we assumed that the pseudoscalar was 7 GeV, which corresponds to $\epsilon_{\mu \tau} = 0.8\%$.  For a lighter pseudoscalar (e.g., 4 GeV), $\epsilon_{\mu \tau}$ is nearly double this value.  In this case, the signal limits can increase by as much as a factor of two.  To first order in $m_{a^0}^2/m_{h^0}^2$, the branching fraction of the Higgs into the pseudoscalar is independent of $m_{a^0}$ and the contour lines in Fig.~\ref{fig: excplot} are unaffected.  Therefore, if the pseudoscalar is near the tau threshold, the experiments are even more sensitive to the Higgs production cross section than indicated in the figure. 

\section{Conclusion}

We have shown that if the Higgs decays into a pair of light pseudoscalars that subsequently decay into taus, then the discovery of the Higgs boson is promising through the subdominant channel where one pseudoscalar decays to a pair of muons.   The Tevatron has a chance of discovering this class of models if CDF and $\DO$ perform combined analyses with the full data sets.  The Tevatron can begin to recover the parameter space that LEP missed with their $h^0\rightarrow 4\tau$ search, which was prematurely stopped at 86\GeV.  Assuming that the only new decay mode of the Higgs boson is into a pair of pseudoscalars, the Tevatron is sensitive to $m_{h^0} \simeq 102\GeV$ with 10 $\fb^{-1}$, and up to $m_{h^0}\simeq 110\GeV$ with 20 $\fb^{-1}$.  When the Tevatron covers this ground, their results, combined with the direct limits from LEP, will effectively establish a lower limit on the Higgs mass regardless of the admixture of Higgs decays into light pseudoscalars or Standard Model fermions.  With a 20 $\fb^{-1}$ cross section, the Tevatron will be sensitive to Higgs bosons up to $m_{h^0} \lsim 150\GeV$.   

At the LHC, this search becomes a method of discovering the Higgs with early data -- potentially with sub-fb${}^{-1}$ data sets.  With an integrated luminosity of $\OO(1\, \fb^{-1})$,  the LHC will be able to recover the missing LEP limits.   Eventually, the LHC will be able to push this branching ratio down substantially, to the 3\% level.  A discovery or even a limit on such a decay mode will be an important step in verifying the field content and symmetry structure of the Higgs potential.

\section*{Acknowledgements}
We are especially grateful to Andy Haas for collaboration throughout the course of this work.  The results of the $\DO$ search for $h^0 \rightarrow a^0 a^0 \rightarrow 2\mu 2\tau$ with 3.7 fb$^{-1}$ have been published \cite{DOnote}.
We would also like to thank Johan Alwall for his assistance with MadGraph/MadEvent, as well as Spencer Chang, Chris Hays, David E. Kaplan,  Aaron Pierce, Philip Schuster, Natalia Toro, and Neil Weiner for helpful discussions.  ML and JGW are supported by the DOE under contract DE-AC03-76SF00515 and partially by the NSF under grant PHY-0244728.  ML is supported by NSF and Soros fellowships.

\end{document}